\documentclass[twocolumn,showpacs,preprintnumbers,amsmath,amssymb]{revtex4}
\usepackage{dcolumn}
\usepackage{bm}
\usepackage[dvipdf]{graphicx}
\DeclareGraphicsExtensions{.jpg,.pdf,.mps,.png,.eps,.ps,.EPS}
\begin{document}
\newcommand{\avg}[1]{\langle{#1}\rangle}
\newcommand{\Avg}[1]{\left\langle{#1}\right\rangle}
\def\be{\begin{equation}}
\def\ee{\end{equation}}
\def\bc{\begin{center}} 
\def\ec{\end{center}}
\def\bea{\begin{eqnarray}}
\def\eea{\end{eqnarray}}

\title{On the flexibility of complex systems}
\author{G. Bianconi$^1$, R. Mulet$^2$}
\affiliation{$^1$The Abdus Salam International Center for Theoretical Physics, Strada Costiera 11, 34014 Trieste, Italy \\
$^2$Henri-Poincar\'e-Chair of Complex Systems and
Department of Theoretical Physics, Physics Faculty, University of 
Havana, La Habana, CP 10400, Cuba}
\date{\today}

\begin{abstract}
Many complex systems satisfy a set of constraints on their degrees of
freedom, and at the same time, they are able to work and adapt to different conditions. 
Here, we describe the emergence of this ability in a simplified model in which the system must satisfy a set of random dense linear constraints. 
 By statistical mechanics techniques, we describe the transition
 between  a  non-flexible system in which the constraints are not fully satisfied, to a flexible system,  in which the constraints can be satisfied in many ways.   This phase transition is described in terms of the appearance of zeros modes in the statistical mechanics problem.
\end{abstract}
\pacs{: } 
\maketitle

Much of the  complexity of biological and social systems is  rooted in  their ability to function in a large  variety of conditions.
At the same time, complex systems must satisfy real constraints. For example, 
the metabolic network to function must, at least, satisfy the steady state equations of the intermediate metabolites production \cite{fba,meta}, and a financial market to be efficient must have a zero excess demand for any distinguishable state of the market \cite{Book}. 
However, to preserve this flexibility,  the constraints can not completely determine
 the state of the system. This will be fully defined by the external
conditions, some optimization principle, and/or the previous history of the system.

In the case of  metabolic networks, all steady states of fluxes
$\{s\}$ are defined by the system ${\bf A \cdot s}-\bf{g}=0$ such that
$0 \leq s_i \leq s_i^{max}$ are intervals that define a convex
manifold $\Omega$. The  matrix $ {\bf A}$ in this problem is  the $M
\times N$ stoichiometric matrix of all the reaction in the network,
with $M$ metabolites and $N$  metabolic fluxes, while the $M$
dimensional vector $\bf{g}$ is different from zero only for
metabolites in input or in output to the network. The real state of
the cell is required to be one of these steady states
\cite{fba,Segre}. In addition, it is usually assumed that the actual
state optimizes the cell growth rate, 
but  this assumption have to be relaxed in various cases
\cite{Segre}. 

This description of the problem is reminiscent of the
Theory of Linear Programming \cite{Chvatal}, 
where one must optimize a linear function
of unknown variables subject to linear constraints. Like in metabolic
networks, the constraints define the space of possible solutions, while the
optimization function, chooses a given one.

A similar problem appears in the study of the Minority Game (MG)
\cite{Book,MG1}. The Minority Game is  a simple model that was
proposed recently to  mimic the dynamics of agents in a market. In one
of the version of this game, the Gran Canonical Minority Game
\cite{GCMG}, the traders in the market, at each time step have access to an external information $\mu=1,\dots, M$ which describes the market state. 
Each agent $i$ is provided with a single strategy indicating which action (buy/sell)  to take given the information $\mu$, i.e.  $a^{\mu}_i=\pm 1$ . There are two types of agents in the markets: $N_s$ speculators that can decide to trade or not to trade depending on their expected success and $N_p$ producers which always play their strategy.
 This game 
has been approached by statistical mechanics \cite{MG1,GCMG} techniques and it
has been proved that the dynamics described above minimizes the Lyapunov function
$H=\sum_{\mu} \left(\sum_{i=1}^{N_s} a^{\mu}_i s_i- B^{\mu} \right)^2 /M$. The soft variable $s_i\in (0,1)$ corresponds to the average use of the strategy of agent $i$ in the stationarity state of the game and  $B^{\mu}=\sum_{i=N_s+1}^{N_s+N_p} a^{\mu}_i$  indicates the information injected by the producers in the market.
Below a critical value of $\alpha=M/N$ the systems is efficient, i.e. it is  such that the expected excess demand $ \sum_{i=1}^{N_s} a^{\mu}_i s_i- B^{\mu} =0$ for each information $\mu=1,\dots,M$  and $H=0$.

It becomes then transparent that a very important question in the study
of complex systems is to know how flexible the system is, i.e.
how large  is  the volume $V$ of the possible solutions as a function
of the parameters of the equations. In the examples described above, it is equivalent to find the number of steady states of the metabolic fluxes in a cell or 
the number of agent actions which clear a financial market.

Technically the problem can be cast in terms of finding the number of  solutions of a linear system of equations when the variables are constrained in a connected finite volume $\Omega$.
Unfortunately,
it is well known that the enumeration of all the vertex of the convex set containing 
these solutions is a $\#P$ complete problem \cite{NumberP}. 
In a more general context, the calculation of the volume of a complex
polyhedron in the $N$ dimensional space, in the so call
$H$-representation (see for example \cite{Fukuda}),  is expressed by a set of $M$ linear inequalities ${\bf A \cdot s} \geq 0$, where   $ {\bf A}$ is again a $M\times N$ matrix.

Similar constraint satisfaction problems have been successfully treated by
statistical mechanics techniques. 
In the case of boolean variables it is known that there is a  transition between  satisfiable to unsatisfiable (SAT-UNSAT) instances as a function of the ratio $\alpha$ between the number of equations $M$ and variables $N$.  Through statistical mechanics methods  it has been proven \cite{Coloring,KSAT2,KSAT3} that in  many {\it NP}-complete problems this SAT-UNSAT transition is preceded by a dynamical transition identified with the absence of a replica  symmetric solution and with the clusterization of the phase space.  
On the contrary if  the problem is defined on continuous variables, no general approach describing the transition is known. Usually, the solution of the volume of the satisfied space is found by imposing a spherical constraint on the variables \cite{Gardner}.
However, as it has been recently shown in the Minority Game context \cite{Galla},
in many situations the spherical constrain generates a different
scenario, and in addition, it is often not justified.

In this paper, using statistical mechanics methods, we describe, as a function of $\alpha=M/N$, the transition between satisfiable and unsatisfiable dense linear systems  of equations
defined  on   continuous variables. In fact it is intuitively clear that 
there is a specific value of $\alpha_c$ such that for $\alpha>\alpha_c$ the system is not able to satisfy
all the constraints while for $\alpha<\alpha_c$ the system becomes satisfiable
for different values of the variables. Here we determine  the
volume of the space of these solutions  and the average values of the variables in term of their 
admitted values, and the statistical properties of the equations. 
 
 For simplicity, we will consider $N$ variables and $M$ equations with random coefficients $\xi_i^{\mu}$ chosen from a distribution $P(\xi)$  with
zero average $\langle \xi\rangle=0$ and  with variance  $\langle
{\xi}^2\rangle=\frac{\sigma^2}{N}$. The system which we would like to be satisfied is
\be
\sum_i \xi^{\mu}_i s_i -g^{\mu}=0 \mbox{\hspace*{1cm}} \forall \mu=1,\alpha M.
\label{eqsystemeq}
\ee
 where $g^{\mu}$ represents the inhomogeneities of the equations, i.e. for example the input or output fluxes in metabolic networks.
 The continuous variables  variables $\{ s_i\}$  would be defined over a set $\Omega =\omega^N$ with $s_i\in \omega$ and $|\omega|=L$.  In the case of convex space $\Omega$ we would like  to address which are the typical features  of a generic problem with $N$ variables and $M$ equations for a generic matrix $((\xi_i^{\mu}))$.

Using a Gaussian representation for the delta functions $\prod_{\mu}\delta(\sum_i \xi^{\mu}_i s_i-g^{\mu})$ imposed by the hard constrains (\ref{eqsystemeq}) 
 we get the  following definition of the unormalized volume of solutions
\be
\tilde{V}(\beta)=\int_{\Omega}\prod_i ds_i \prod_{\mu} \frac{e^{-\beta\left(\sum_i \xi^{\mu}_i s_i -g^{\mu}\right)^2}}{\sqrt{ \pi/\beta}} .
\label{eqdefvolumen-beta}
\ee
to be taken in the limit $\beta\rightarrow \infty$. Consequently we have introduced a Gaussian measure  $f_0=
\prod_{\mu} {e^{-\beta\left(\sum_i \xi^{\mu}_i s_i -g^{\mu}\right)^2}}/(( \pi/\beta)^{M/2}\tilde{V}(\beta)$ in the space of the variables $\Omega$.

This kind of problem, within the formalism of statistical physics, can be approached defining a Hamiltonian $H$ to be  minimized when the system is solved. For our case of interest $H$ is given by
\be
H=\sum_{\mu} \left( \sum_i \xi^{\mu}_i s_i -g^{\mu}\right)^2
\label{eqdefenergy}
\ee
\noindent
 which is equivalent to the one defined in the study of the Minority Game \cite{Book}.

For such a system the partition function reads
\be
Z=\int_{\Omega}  \prod_i ds_i \exp \left[{-\beta \sum_{\mu} \left(\sum_i \xi_i^\mu s_i-g^{\mu}\right)^2}\right].
\label{eqpartfunction}
\ee
and  the entropy is given by 
\be
S=-<\log(f_0)>=\beta E+\frac{M}{2}\log\left(\frac{\pi}{\beta}\right)+\log(\tilde{V})
\label{eqentropy}
\ee

Thus our goal would be to evaluate the free energy of this system mediated over the quenched disorder
$\frac{-1}{\beta}\langle\log(Z)\rangle=F=E-TS$, being $E$ the energy  and $S$ the entropy of the system.
Obviously, from our Hamiltonian (\ref{eqdefenergy}), we must have, 
 $E=0$ for satisfiable systems and $E\neq 0$ for unsatisfiable systems. 

The calculation of $\langle \log(Z)\rangle $  proceeds using the usual replica trick
$\langle \log(Z)\rangle =\lim_{n\rightarrow 0} \frac{\langle Z^n
  \rangle -1}{n}$ 
in which $Z^n$ indicates the $n$ times replicated partition function with the
same quenched noise. Then, using a  Replica Symmetric ansatz
\cite{MPV}, and on similar lines 
to the analogous Hamiltonian of the Minority Game \cite{Book,MG1}  we
obtain a result, which, extended to the limiting case $L\rightarrow
\infty$, $\beta\rightarrow \infty$ (which is the Algebra limit) provides an energy $E/N=\frac{1}{2}\langle g^2\rangle (\alpha-1)+\frac{1}{2 \beta}$
and an entropy $S/N=-\frac{\alpha}{2}\log \alpha +\frac{\alpha-1}{2}\log\left(\alpha-1\right)+\frac{1}{2}\log\left(\frac{\pi e^2}{\sigma^2\beta}\right)$.
This solution is wrong for $\alpha<1$. In fact the energy becomes negative,
contradicting equation (\ref{eqdefenergy}). Moreover for $\alpha<1$  
the entropy  must acquire a term of the order of $N\log(L)$ that here is not present. This signals that the standard RS ansatz in the $ L,\beta \rightarrow \infty$  fails for $\alpha<1$.

In order to solve this problem, which  arises also in the finite $L$ case, we gain insight from the known algebra solution of the case $L\rightarrow \infty$. 
In this case the  linear system of equations  is  solved first fixing, 
arbitrarily, $N-M$ variables and then solving respect to the others. Consequently, to solve the problem at finite $L$ we assume that although all the variables are equivalent, below the satisfiable transition an arbitrary set of $(1-m)N$ variables is free to assume any value $s_i\in \omega $ without fluctuations while  the other $mN$ variables can still satisfy the linear problem.
Then, a new  variational parameter $m$ is introduced in  the free energy and fixed through the saddle point equations.
The constrained variables $i=1,\dots m N$  would have an overlap  ${Q}_{a,b}=<s^a s^b>$ and conjugated variables $\hat{Q}_{a,b}$  Replica Symmetric,i.e. $Q_{a,b}=Q\delta_{a,b}+q (1-\delta_{a,b})$, $\hat{Q}_{a,b}=\hat{Q}\delta_{a,b}+\hat{q} (1-\delta_{a,b})$
On the contrary  the other $(1-m)N$ variables would have and average $
\sum_{i=mN}^N s_i^a=(1-m) N S^a $, and an overlap
$\sum_{i=mN}^N s_i^a s_i^b=(1-m) N {S^a S^b} .$ 
Where $S^a=S$ and $\hat{S}^a=\hat{S}$, keeping the RS scenario.

Using this ansatz the free energy reads,
\begin{widetext}
\bea
\beta f&=& -(1-m) \hat{S}{S}-m ( Q \hat{Q}-q \hat{q} )+\frac{\alpha}{2} \biggl[
  \frac{\frac{g^2}{\sigma^2 m}+q+\frac{1-m}{m} S^2}{Q-q+\frac{1}{2 \beta \sigma^2 m}}+\ln {\bigg(Q-q+\frac{1}{2 \beta \sigma^2 m} \bigg) }+\ln (2\beta \sigma^2 m)\biggr] \nonumber \\ &&  -m \int_{-\infty}^{\infty} \frac{dy}{\sqrt{2 \pi}} e^{-y^2/2} \ln H_L(\hat{q},\hat{Q}) -(1-m) \ln L-(1-m) \ln\left(I(\hat{S})\right)
\label{fe}
\eea
\end{widetext}
 and the saddle point equations become
\bea
m \hat{q}&=&-\frac{\alpha}{2} \frac{\frac{g^2}{\sigma^2 m}+q+\frac{1-m}{m} S^2}{\big(Q-q+\frac{1}{2 \beta \sigma^2 m}\big)^2} \nonumber \\
m \left(\hat{Q}-\hat{q}\right) &=&\frac{\alpha}{2}
\frac{1}{Q-q+\frac{1}{2 \beta \sigma^2 m}} \nonumber \\
Q&=& \langle \langle s^2 \rangle \rangle  \label{eqsaddlepoint} \\
Q-q&=&\frac{1}{\sqrt{-2\hat{q}}} \langle \langle y s \rangle \rangle \nonumber\\
\int_{-\infty}^{\infty} Dy \ln H_L(\hat{Q},\hat{q})&=&\ln\left( L
  I(\hat{S}) \right)+ S \hat{S}-(\hat{Q}-\hat{q})S^2\nonumber 
\eea
to be considered together with the  two equations
\bea
\frac{\partial  f}{\partial S}=0 &\mbox{i.e.} \ &\hat{S}=2S\big(\hat{Q}-\hat{q}\big)\nonumber \\
\frac{\partial  f}{\partial \hat{S}}=0 & \mbox{i.e.}\  &  S=\avg{s}_{\hat{S}}
\label{SSh}
\eea
 with $H_L(\hat{Q},\hat{q})=\int_{\omega} ds e^{W(s,y)}$ , $W(s,y)=-(\hat{Q}-\hat{q})  s^2 +\sqrt{-2 \hat{q}} y s$, $I(\hat{S})=\int_{\omega} ds e^{-\hat{S}s}/L$ $\avg{s}_{\hat{S}}=\int_{\omega} ds s e^{-\hat{S}s}/\int_{\omega} ds  e^{-\hat{S}s}$ and finally $g^2=\sum_{\mu} (g^{\mu})^2/M$. 

If the interval $\omega$ is symmetric respect to zero, the solution $\hat{S}=S=0$ is always allowed. When $\omega$ is not symmetric respect to zero it is always possible to  translate the variables and to define the problem on a  symmetric interval around zero.  In this translation the variables  change following $s_i\rightarrow s_i-x_0$ and the constrain following  $g^{\mu}\rightarrow g^{\mu}+\sum_ia^{\mu}_i x_0$ ($g^2\rightarrow g^2+\sigma^2 x_0^2$) with $x_0$ indicating the center of mass of the interval $\omega$.
In the special case of  the algebra limit, $L,\beta \rightarrow \infty$,  this solution for $\alpha < 1$, predicts: the free energy
$\beta f= -\frac{\alpha}{2} \log\left(\frac{\pi e}{\sigma^2 \beta}\right)-(1-\alpha)\ln L
$, the  energy $ E/N=\frac{\alpha}{2\beta}$ and the entropy $S/N= \frac{\alpha}{2} \log\left(\frac{\pi e^2}{\sigma^2\alpha \beta}\right)+(1-\alpha)\ln L$. Consequently for $\alpha<1$ the system is satisfiable in the limit $\beta\rightarrow \infty$, in fact 
$E=\frac{\alpha}{2\beta}\rightarrow 0$  and the volume of solutions
 $\tilde{V}$ given by $e^{\langle \log (\tilde{V})\rangle}=L^{N-M} \left(\frac{e}{\alpha \sigma^2}\right)^{M/2}$ as it should since $\left(\frac{\alpha\sigma^2}{e}\right)^{M/2}$ is the determinant of a $M \times M$ matrix of  coefficients $\xi^{\mu}_i$ with $\avg{\xi}=0$ and variance $\avg{\xi^2}=\sigma^2/N$ \cite{Mehta}.

In the case in  which   $\omega$ is bounded  and $L$ finite, from 
(\ref{eqentropy})  and (\ref{fe}),
we have for the volume of solutions, 

\bea
\frac{1}{N}\langle \ln \tilde{V}\rangle=& \ln L+m(Q\hat{Q}-q\hat{q})-\frac{\alpha}{2} \frac{\frac{g^2}{\sigma^2 m}+q}{Q-q+\frac{1}{2 \beta \sigma^2 m}}\nonumber \\ &-\frac{\alpha}{2}\ln {\left[{2\pi m \sigma^2} \bigg(Q-q+\frac{1}{2 \beta \sigma^2 m} \bigg) \right]} . \label{eq:volume}
 \eea

\noindent with the saddle point equations defining the value of the
variational parameters.
In this case, the saddle point equations
(\ref{eqsaddlepoint}) must be solved numerically and the validity of our
ansatz may be tested with simulations.

On the other hand, to estimate the average values of the variables one
 must consider that while  the equations
 $(\ref{eqsaddlepoint}),(\ref{SSh})$ allow for a solution with $m<1$
 the system is satisfiable and a series of $1-m$ zeros modes are
 present in the solution. These zero modes are associated to the
 freedom in which the $(1-m) N$ variables can be chosen.
 In this case, as a fraction $(1-m)$ of free variables can  assume any
 value in the set $\omega$  the average value of the variables
 $\langle s \rangle$ for each realization of the system is not fixed
 by the value of $\alpha$.
Consequently the average value of the variables $\langle s\rangle $ and the second moment $\langle s^2\rangle$ for a single realization of the system are described as  
$\overline{s} =m \langle\langle s \rangle\rangle_W+(1-m) s_0 $ and 
$\overline{s^2} =m \langle\langle s^2 \rangle\rangle_W+(1-m) s_0^2$
 where  $\langle \langle \dots\rangle \rangle_W$ indicates the 
average of the constraint variables on the effective potential
$W(s,y)$ and  $s_0$ indicates the average value of the free variables in the
considered realization.

\begin{figure}
\includegraphics[width=75mm,height=55mm]{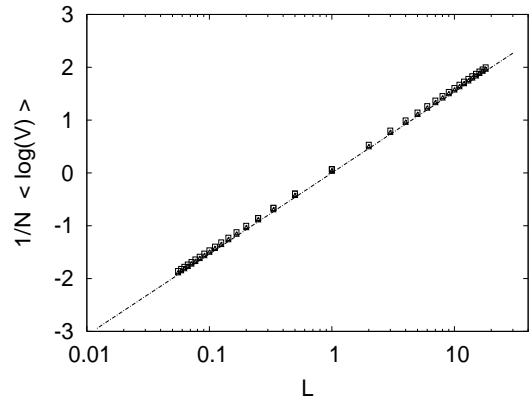}
\caption{The volume of solutions of a linear system of  equations
  $\alpha=1/3$. 
The continuous line corresponds to the theoretical prediction, 
and the symbols to the simulations with  $N=6$ ( $\boxdot$), $N=9$ ($\triangleleft $).}
\label{figvolume}
\end{figure}
We note here that the  observable volume of solutions $V$ differs from $\tilde{V}$ and is defined as 
\be
V(\beta)=\int_{\Omega}\prod_i ds_i \prod_{\mu} \frac{e^{-\beta\left(\sum_i \xi^{\mu}_i s_i -g_{\mu}\right)^2}}{\sqrt{ \pi/\beta}} \det{{\cal R}(mN \times mN)}  \nonumber
\ee 
where $\cal R$ is a square random matrix of   $mN$ elements and $m$ is a the fraction of constraint variables to be determined by saddle point equations.
The direct calculation of $\avg{\log(V)}$  can be approximated very well 
with 
\be
\avg{\log(V)}\approx\avg{\log(\tilde{V})}+\log[\det{{\cal R}(mN \times mN)}]
\label{approx}
\ee
with $m$ given by the solutions of (\ref{eqsaddlepoint}) as we checked directly.

To test the  calculations we compute numerically the volume of
 solutions of randomly generated linear equations 
and compared it with our theoretical predictions. 
As already mentioned in the introduction, 
finding  the volume of a polytope is a
$\#P$-complete problem, therefore the  direct calculation of the
volume of solutions is restricted, in practice, to  problems with a
small number of equations and variables. 
To simplify the computations, the elements of the matrix $A$ were
taken as $\pm 1$ with probability $\frac{1}{2}$ and $g^{\mu}=0$. 
The variables were constrained to be in $(-L/2,L/2)$

The simulations were done using {\it vinci} \cite{Fukuda} 
and the related {\it lrs} code \cite{lrs}, which shows reasonable 
performance in time for dense systems of no more
than $M=4$ equations and $N=12$ variables. Moreover, one must keep in
mind that the volume computed by {\it lrs} in these not full dimensional
polytopes is the projection on the lexicographically smallest
coordinate subspace\cite{lrs}. For the matrices we used, this is equivalent to rescale
the volume by a factor $N^{\frac{1}{2}}$ where $N$ is the number of variables.

The comparison between the numerical results and our analytical
predictions appear in  figure \ref{figvolume} where we show the
direct average over 100 instances of systems of equations with
$N=6$ and $9$ variables and $\alpha=1/3$ as a function of $L$. The
analytical results were obtained trough the substitution in equations
$(\ref{approx})$ and $(\ref{eq:volume})$
of the solutions of the system of equations
$(\ref{eqsaddlepoint})$ with $S=\hat{S}=0$. As can be immediately
seen, the agreement between our simulation and the theoretical line is
very good. This, despite the fact that our predictions are valid in
the thermodynamic limit and the simulations may be performed only for
very small systems.

In conclusion, we have described the emergence of flexibility in a dense linear systems of equations 
of $N$ real variable and $M$ equations.
This flexibility results from the appearance, for  $\alpha<\alpha_c$,
of an exponentially large number of states that satisfy the constraints imposed to the system.
Our results prove also  that
for  $\alpha=M/N>\alpha_c$, the standard RS
scenario holds and the system is not satisfiable, while for  $\alpha < \alpha_c$  an exponential  number
of solutions are present and the symmetry of permutation of the variables
is broken. In this case, a  fraction $m$ of the variables  must be
considered free while the rest are fixed by the constraints. 
Our analytical predictions were tested  computing numerically
 the volume of random polytopes derived from dense system of linear equations showing that the derivation presented provides the value of 
the caracteristic volume of a random polytope. 
The problem addressed in this paper is stylized and general, 
extensions of this work would  
include both the study of the same model on diluted 
graphs (with different degree distributions) in order to better describe
 the metabolic networks and a detailed analysis  
of the Minority Game. In both cases we expect that a  scenario similar to 
the one discussed above
 holds. Work in these directions is in progress.

G. B. acknowledge  support by the European Community's Human Potential Program
COMPLEX-MARKETS. R. Mulet thanks the International Centre for
Theoretical Physics, ICTP, for the hospitality.


\begin{thebibliography}{}
\bibitem{fba} J. S. Edbitemwards and B. O. Palsson, PNAS {\bf 97} 5528 (2000).
\bibitem{meta} B. Palsson, {\it Nature Biot.} {\bf 18}, 1147 (2000).
\bibitem{Book}
D. Challet, M. Marsili and Y.-C. Zhang, {\it Minority Game} (Oxford University Press,2005).
\bibitem{Segre} D. Segre', D. Vitkup and G. M. Church, PNAS {\bf 99},
  15112 (2002).
\bibitem{Chvatal} V. Chvatal, {\it Linear Programming} (New York, NY,
  Freemanm 1983)
\bibitem{MG1} D. Challet, M. Marsili and R. Zecchina, Phys. Rev. Lett.
{\bf 84} 1824, (2000).
\bibitem{GCMG} D. Challet and M. Marsili, {\it Phys. Rev. E} {\bf 68}, 036132 (2003). 
\bibitem{nota} Properly speacking  the Gran Canonical Minority Game has an additional parameter $\epsilon$ with we will consider here  to be zero ($\epsilon=0$).
\bibitem{NumberP} M.E. Dyer and A. M. Frieze, SIAM J. of Comput. {\bf17} 967 (1988)
\bibitem{Fukuda} B. Bueler and A. Enge and K. Fukuda, {\it In
    Proceedings of the conference `Computational Geometry' }, Munster
  (Germany), March 1996.
\bibitem{KSAT2} M. M\'ezard, G. Parisi and R. Zecchina, Science {\bf 297}, 812 (2002).
\bibitem{Coloring} R. Mulet, A. Pagnani, M. Weigt and R. Zecchina, Phys. Rev. Lett {\bf 89} 268701 (2002).
\bibitem{KSAT3}  A. Braunstein, M. M\'ezard and  R. Zecchina,  Random
  Structures and Algorithms {\bf 27}, 201 (2005)
\bibitem{Gardner} E. Gardner J. Phys. A {\bf 21},257 (1988).
\bibitem{Galla} T. Galla and D. Sherrington, cond-mat/0508413 (2005).
\bibitem{MPV} M. M\'ezard, G. Parisi, M. A. Virasoro, {\it Spin Glasses and Beyond}, (World Scientific,Singapore,1987).
\bibitem{Mehta} M. L. Mehta, {\it Random matrices} (Academic Press, San Diego, 2ed. 1991).
\bibitem{lrs} D. Avis, {\it  Polytopes - Combinatorics and Computation}, G. Kalai and G. Ziegler eds., Birkhauser-Verlag, DMV Seminar Band 29, pp. 177-198 (2000). 
\end{thebibliography}
\end{document}